\documentclass[letterpaper,11pt]{article}

\usepackage{fontawesome}

\usepackage{xcolor}
\usepackage{xurl}
\usepackage{tabularx} 
\usepackage{tikz}
\usepackage{amsmath}
\usepackage{amsthm}
\usepackage{dsfont}
\usepackage{enumitem}
\usepackage{amssymb}
\usepackage{subfigure}
\usepackage{tablefootnote}

\newcolumntype{L}[1]{>{\raggedright\arraybackslash}p{#1}}

\usepackage{array}
\newcommand{\PreserveBackslash}[1]{\let\temp=\\#1\let\\=\temp}
\newcolumntype{C}[1]{>{\PreserveBackslash\centering}p{#1}}
\newcolumntype{R}[1]{>{\PreserveBackslash\raggedleft}p{#1}}
\newcolumntype{L}[1]{>{\PreserveBackslash\raggedright}p{#1}}

\usepackage{booktabs}

\makeatletter
\newtheorem*{rep@theorem}{\rep@title}
\newcommand{\newreptheorem}[2]{%
\newenvironment{rep#1}[1]{%
 \def\rep@title{#2 \ref{##1}}%
 \begin{rep@theorem}}%
 {\end{rep@theorem}}}
\makeatother

\newtheorem{definition}{Definition}

\newtheorem{lemma}{Lemma}
\newtheorem{theorem}{Theorem}

\newreptheorem{theorem}{Theorem}
\newreptheorem{lemma}{Lemma}
\newreptheorem{proposition}{Proposition}

\definecolor{DarkGreen}{rgb}{0.075,0.375,0.075}
\definecolor{DarkRed}{rgb}{0.5,0.1,0.1}
\definecolor{DarkBlue}{rgb}{0.1,0.1,0.5}
\definecolor{Gray}{rgb}{0.2,0.2,0.2}

\definecolor{myblue}{rgb}{0.23299120924703914, 0.639586552066035, 0.9260706093977744}
\definecolor{myred}{rgb}{0.9677975592919913, 0.44127456009157356, 0.5358103155058701}
\definecolor{mygreen}{rgb}{0.3126890019504329, 0.6928754610296064, 0.1923704830330379}
\definecolor{myorange}{rgb}{0.9333333333333333, 0.5215686274509804, 0.2901960784313726}

\usepackage{xspace}
\usepackage[bottom]{footmisc}
\usepackage{soul}

\usepackage[square]{natbib}

\renewcommand{\deg}{\mathrm{deg}}

\usepackage{pifont}

\usepackage[small]{caption}
\usepackage[pdftex]{hyperref}
\hypersetup{
    unicode=false,          
    pdftoolbar=true,        
    pdfmenubar=true,        
    pdffitwindow=false,      
    pdfnewwindow=true,      
    colorlinks=true,       
    linkcolor=DarkBlue,          
    citecolor=DarkGreen,        
    filecolor=DarkRed,      
    urlcolor=DarkBlue,          
    pdftitle={},
    pdfauthor={},
}

\usepackage[margin=1.1in]{geometry}

\usepackage{authblk}
\stepcounter{footnote}
\addtocounter{footnote}{-1}
\author[1]{Dorothee Sigg}
\author[1]{Moritz Hardt}
\author[1,2]{Celestine Mendler-Dünner}
\affil[1]{Max-Planck Institute for Intelligent Systems, Tübingen and Tübingen AI Center}
\affil[2]{ELLIS Institute, Tübingen}
\date{}                 
\title{Decline Now: A Combinatorial Model for Algorithmic \\Collective Action}
\date{}

\begin{document}
\maketitle

\begin{abstract}
Drivers on food delivery platforms often run a loss on low-paying orders. In response, workers on DoorDash started a campaign, \#DeclineNow, to purposefully decline orders below a certain pay threshold. For each declined order, the platform returns the request to other available drivers with slightly increased pay. While contributing to overall pay increase the implementation of the strategy comes with the risk of missing out on orders for each individual driver.
In this work, we propose a first combinatorial model to study the strategic interaction between workers and the platform. Within our model, we formalize key quantities such as the average worker benefit of the strategy, the benefit of freeriding, as well as the benefit of participation. We extend our theoretical results with simulations. Our key insights show that the average worker gain of the strategy is always positive, while the benefit of participation is positive only for small degrees of labor oversupply. Beyond this point, the utility of participants decreases faster with increasing degree of oversupply, compared to the utility of non-participants. Our work highlights the significance of labor supply levels for the effectiveness of collective action on gig platforms. We suggest organizing in shifts as a means to reduce oversupply and empower collectives. 
\end{abstract}

\section{Introduction}

The emergence of gig platforms has changed the way in which people work. Acting as mediaries, they facilitate matching and direct transactions between customers and labor force. Gig workers are treated as independent contractors, and elaborate technology is used to orchestrate on-demand services. Platforms hire psychologists, social scientists and data scientists to implement a machinery of algorithms that control, profile, shape and discipline workers’ behaviour \citep{dubal2020brief}. A huge amount of data is collected and used for scoring, classification, ranking and prediction \citep{woodcock2019gig, vallas2020platforms, liu2022machine}. On top of that, gig platforms use ``a combination of an oversupply of workers and financial incentives [...] to still have some influence on when their workers work" \citep{woodcock2019gig}.

Gig workers on the other hand find themselves in a workplace marked by ``low pay, precarity, stressful and dangerous working conditions, one-sided contracts and a lack of employment protection" \citep{woodcock2019gig}. Wages are not transparent – ``calculated with ever-changing formulas using granular data on location, individual behavior, demand, supply, and other factors" \citep{dubal2023algorithmic}. Over the past years, workers have taken up numerous strategies and initiatives that aim to improve labor conditions across the gig economy~\citep{robinson2017making, savage2020becoming, mohlmann2017hands, jarrahi2019algorithmic}. Especially ride hailing and food delivery platforms have been at the center of ongoing political struggle~\citep{dubal2022economic} and labor activism \citep{wood2019platform, wells2021just}.

\subsection{Our work}

DoorDash is presently the largest food delivery service in the United States with a market share of approximately two thirds of food deliveries. In 2019, drivers on DoorDash organized a campaign to collectively raise earnings per ride. Dubbed \#DeclineNow, participating workers would decline rides below a minimum pay threshold.

Inspired by \#DeclineNow, we propose a combinatorial model of worker collective action. In our model, a pool of $N$ drivers serves the requested deliveries, one per time-step. Drivers participating in the collective decline orders below a certain threshold pay. Non-participating drivers accept the request at any price. The platform returns any declined order to a random available worker at a slightly increased price, until a worker accepts the order.

Through theoretical analysis and extensive simulations, we answer fundamental questions about the feasibility and benefits of collective action in our model. We quantify the average increase in earnings across all workers and show that it is always positive.  Moreover, the benefit of participating in the collective is positive in the case of labor undersupply, where workers are fewer than orders. In this case, there is no \emph{freeriding}, i.e., returns to non-participating workers. In the case of labor oversupply, however, freeriding has positive returns and the benefit of participating may no longer be positive. We provide sufficient conditions for collective action to be self-incentivized, relating collective size and supply condition. Numerical simulations extend our observations, specifically, to settings that are beyond our theoretical analysis.

Our work complements a recent study of collective action on platforms that deploy machine learning algorithms, such as a resume screening model on a freelance platform~\citep{hardt2023algorithmic}. The authors of the study concluded that even small collectives can have significant power over a platform that allocates on the basis of a predictive model. Here, we study a combinatorial model of a platform that acts according to a fixed mechanism rather than a learned rule. The rigidity of the allocation rule naturally gives the collective fewer levers to turn. As a result, the return to collective action is generally smaller. 

Nonetheless, we identify market conditions under which collectives can be effective. Based on these insights our results suggest important practical avenues for improvement. We demonstrate that by organizing into shifts, for example, drivers can significantly reduce worker oversupply and hence amplify their leverage.

\paragraph{Outlook.} In Section~\ref{sec:background} we start by providing historical background on the emergence of  algorithmic collective action on gig platforms. We collect empirical evidence of 30 cases of collective action, demonstrating how platform participants aim to use technology-specific levers to contest algorithmic systems. Then, inspired by the \#DeclineNow case, we propose a combinatorial model for algorithmic collective action in Section~\ref{sec:model}, instantiated with empirical evidence about worker pay and model parameters as detailed in Section~\ref{sec:paraminstantiation}. 
Through theoretical analysis and simulations we gain valuable insights into the key factors driving the success of the collective. Results are presented in Section~\ref{sec:results}. Building on our insights, in Section~\ref{sec:workinginshifts} we propose a strategy to work in shifts for the collective to achieve more favorable market conditions. Section~\ref{sec:discussion} concludes our work with a broader discussion.

\section{From traditional strikes to algorithmic collective action}
\label{sec:background}

The origins of the gig economy, as we know it today, go back to approximately 2009 when Uber was founded. Protests emerged a couple of years later \citep{dubal2020brief} reaching a first peak with the 2016 strikes in London and Turin against Deliveroo and Foodora \citep{wood2019platform}. Strikes, along with demonstrations, open letters and collective bargaining offer a form of protest that has been used by workers for decades. 
The archive on collective action in tech (CAIT)\footnote{\url{https://collectiveaction.tech/}, accessed on 2023-08-18.} comprises over 500 accounts of strikes, protests, and open letters among others. A similar data collection from media reports is accomplished by the Leeds Index of Platform Labour Protest,\footnote{\url{https://leeds-index.co.uk/}, accessed on 2023-08-28.} where \citet{bessa2022global} identify pay as the most important cause for workers' protests.

However, these traditional forms of collective action cannot always be transferred equally to the new, algorithm-driven work environment of the gig economy – it might differ both in terms of organization and effect. For instance, one issue with strikes is that other gig workers might be incentivized to work if others don't – ``you could make a lot more money than you normally would by being the only Uber driver" \citep{magesan2019uberdrivers}. Instead, the digital nature of gig work calls for and opens up new, technology-specific collective levers. While strategies on an individual level exist (e.g. exploiting a bug in an algorithm), a group oftentimes has much more power as it controls more data. We refer to technology-specific efforts of collectives in algorithm-managed work spaces as \emph{algorithmic collective action}.

\subsection{Empirical evidence of algorithmic collective action}
There are many documented cases of algorithmic collective action efforts in algorithmically managed workspaces across various sectors. By searching forums, Facebook groups, Reddit threads, newspaper articles and scientific literature we collected 30 cases of algorithmic collective action and grouped them by the applied mechanism. All cases are listed in Table~\ref{tableofcases}, which we offer as a  living archive on Github\footnote{\url{https://github.com/dorothee-sigg/collective-action}} so it can be expanded by the community with future cases. We discuss three technology-specific levers applied throughout cases:

\begin{table}[h!]
\scriptsize
    \centering
    \caption{Documented cases of algorithmic collective action. Category which describes the applied mechanism for each case (collective information (CI) sharing or collective strategizing (CS)).}
    \label{tableofcases}
\begin{tabular}{|@{\hskip.05cm}L{2.7cm}@{\hskip.05cm}|@{\hskip.05cm}L{7.5cm}@{\hskip.05cm}|@{\hskip.05cm}C{1.3cm}@{\hskip.05cm}|@{\hskip.05cm}L{4cm}@{\hskip.05cm}|}
\hline
\textbf{case} &
  \textbf{description} &
  \textbf{category} &
  \textbf{source} \\ \hline
  
Turkerview, Turcopticon\tablefootnote{\url{https://turkerview.com/}, \url{https://turkopticon.net/}}&
  forums for collecting/averaging ratings on requester, sharing individual strategies &
  CI &
  \cite{salehi2015we}  \\ \hline
  
Uber, surge verification &
  drivers notify each other to verify accuracy and profitability of surge areas &
  CI &
  \cite{dubal2023algorithmic} \\ \hline
  
Truck drivers &
  electronic logging devices: brute-force destruction, sharing log-in information, slow rolls, collaboration between drivers and truck stops &
  CI / CS &
  \cite{levy2023data} \\ \hline
  
Instagram influencers & understanding of algorithm by reporting on experiments  &
  CI &
  \cite{omeara2019weapons} \\ \hline
  
Upwork workers & forums to jointly making sense of the algorithms logic on e.g. account-visibility, customer ratings & CI & \cite{jarrahi2019algorithmic}\\ \hline

WeChat groups &
  sharing individual strategies (temporary subsidies, newcomer help), joint sensemaking of the algorithm &
  CI &
  \cite{sun2019your, yu2022emergence} \\ \hline
  
Workers Info Exchange\tablefootnote{\url{https://www.workerinfoexchange.org/}} &
  help workers to make a data request from the gig platform, research with collective data &
  CI &
  \cite{safak2021managed} \\ \hline
  
Driver's Seat Cooperative &
  driver-owned data collection/analytics service, providing performance data for drivers &
  CI &
  \url{https://driversseat.co/} \\ \hline
  
WeClock &
  self-tracking app for workers, collected data used for campaigns &
  CI &
  \url{https://weclock.it/} \\ \hline

Postmates \#BlitzUp & one-day protest; only accepting ``Blitz" offers & CS & \cite{postmatesblitzup} \\ \hline

Amazon Flex workers & phones in trees + syncing them with workers' phones, advantage over location-dependent offers & CS & \cite{soper2020treephones} \\ \hline

Uber surge club & log-out, trigger supply-/demand-based surge pricing & CS & \cite{mohlmann2017hands, robinson2017making} \\ \hline

Youtube & campaign to extensively use community polls  & CS & \cite{thespiffingbrit} \\ \hline

Spotify, song length & band earned money by asking their fans to listen to songs that consist of 31 seconds of silence & CS & \cite{welch2015spotify} \\ \hline

Spotify, song ranking & singer asked fans to listen to song repeatedly, create playlists, use multiple accounts to hack ranking & CS & \cite{marasciulo2022spotifyanitta} \\ \hline

DoorDash, \#DeclineNow & drivers reject low-paying orders, increased pay for fellow driver & CS & \cite{bloomberg2021doordash, vice2021doordash} \\ \hline

FoggySight & user privacy in facial lookup settings, ambiguous images to obfuscate another user’s identity & CS & \cite{evtimov2020foggysight} \\ \hline

AdNauseam & browser extension randomly clicks on ads, protecting own and other users' privacy as a side effect & CS & \cite{brunton2015obfuscation} \\ \hline

Instacart, 22 cent Tips & tipping 22 cent in app as a signal that more tips will be paid in cash at door & CS & \cite{eidelson2019instacart22, instacart22} \\ \hline

Instacart, \#DeleteInstacart & workers asked customers to delete the app & CS & \cite{deleteinstacart, guardian2021deleteinstacart} \\ \hline

Restaurant-driver cooperation & drivers cooperating with restaurants, placing artificial orders to meet bonus pay requirements & CS & \cite{yu2022emergence, sun2019your} \\ \hline

Instagram, engagement pods & peer community to mutually engage with each others content & CS & \cite{omeara2019weapons} \\ \hline

Upwork, evaluation metric & workers are asking clients to give them specific ratings to enhance their scores & CS & \cite{jarrahi2019algorithmic, rahman2019invisible} \\ \hline

Upwork, off-platform & workers and customers move work off-platform & CS & \cite{jarrahi2019algorithmic, rahman2019invisible} \\ \hline

Twitter, screenshots & posting screenshots of other posts to avoid a forced connection & CS & \cite{burrell2019twitter} \\ \hline

Twitter, voldemorting & not mentioning words or names (instead inventing new names) in order to avoid a forced connection & CS & \cite{burrell2019twitter} \\ \hline

Facebook, remote geotagging & collective remote check-in to obfuscate geolocation data allegedly monitored by police & CS & \cite{baik2020geotagging} \\ \hline

Google bombs & heavy linking of terms or websites for higher rank & CS & e.g. \cite{gillespie2019algorithmically} \\ \hline

Waze, traffic routing & individuals creating bots to influence crowd-sourced navigation systems & CS & \cite{sinai2014exploiting} \\ \hline

Fanfiction writers & produce fake stories to poison data collection services & CS & \cite{fanfictionwriters} \\ \hline

\end{tabular}
\end{table}

\paragraph{Collective information sharing.} One typical characteristic of digital platforms is a high degree of information asymmetry between platform participants and platform owners. As a result digital platform workers often experience a high degree of arbitrariness in algorithmic decisions~\citep[e.g.,][]{rosenblat2016algorithmic,kellogg2020algorithms,woodcock2019gig}. To counter this information imbalance, workers have started to engage in information sharing strategies of different kind.
Workers join efforts to decipher intransparent and non-public algorithms by sharing their experiences, strategies and report conclusions from small experiments. For example, \citet{jarrahi2019algorithmic} provide detailed account of freelance workers at Upwork who spent a significant amount of time to understand various ratings, timeframes, policies, and procedures of the platform. \citet{sun2019your} describes this process of sense-making among Chinese gig workers as inventing ``labor algorithms" which have labor as input and income as output. We also found cases where workers exchange \emph{individual strategies} and make them accessible for a larger group through online forums, messenger services or social media. For instance, Chinese rideshare and delivery drivers are organized in WeChat groups (popular messenger service in China). They share information about temporary subsidies, equipment assistance, newcomer help, real-time traffic news and help each other in emergencies \citep{sun2019your,yu2022emergence}.

In other cases each worker contributes data which is then aggregated as a group. This might yield otherwise non-accessible insights. For instance, Uber drivers confirm or deny in a group chat among their peers whether a surge area (i.e. an area of allegedly higher prices) turns out to be remunerative  \citep{dubal2023algorithmic}. On a larger scale, an example for data aggregation is from two online forums for Amazon Mechanical Turk workers (AMT), Turkopticon and Turkerview. There, workers can rate requesters (i.e. people/organizations who post tasks on AMT) with regard to pay or fairness of tasks. As an average these ratings provide very helpful information for the community of clickworkers which tasks to chose and which requesters to avoid \citep{salehi2015we,hanrahan2021expertise}. Also, data cooperatives provide a form of data aggregation. Data cooperatives are an emergent subject of academic research \citep{pentland2020data,buhler2023data} and refer to collaborative organizations or platforms where individuals voluntarily contribute their personal data for the benefit of the community. An interesting example from the gig economy is the Driver’s Seat Cooperative,\footnote{\url{https://www.driversseat.co/}} a driver-owned, independent cooperative that provides an app to enable ride-hailing drivers and on-demand delivery drivers to collect work and performance data. The cooperative aims at equipping workers with actionable insights to increase their wages.~\looseness=-1

\paragraph{Collective strategizing around training resources.}
The reliance of machine learning powered platforms on human-generated data during training offers an additional lever for workers to exploit. 
An example case for leveraging data to poison training data comes from a group of fan fiction writers in the U.S. who published irreverent stories online ``to overwhelm and confuse the data-collection services that feed writers’ work into AI technology" \citep{fanfictionwriters}. A different example is data obfuscation to deliberately adding noise to make data more ambiguous or harder to exploit \citep{brunton2015obfuscation}. An interesting, so far only theoretical example for collective data obfuscation is FoggySight \citep{evtimov2020foggysight}, an anti-facial recognition technology which aims to enhance privacy in facial lookup settings.

\paragraph{Collective strategizing at deployment time.} There are cases where participants agree upon and execute a joint strategy at deployment time, sometimes in combination with training time strategies. However, in contrast to the latter, deployment-time strategies can also be applied to rigid, combinatorial allocation algorithms. 
An example of users behaving strategically in response to an algorithm is the case of Uber drivers which collectively log out to purposefully manipulate the prices to earn more. By collectively logging out of the app they create an artificial shortage of drivers to trigger surge pricing that they can take advantage of after logging back in \cite{uberconfessions2014,robinson2017making,mohlmann2017hands,uberreagan2019}. A different example are the screenshotting and voldemorting strategies on Twitter. Screenshotting means posting a screenshot of another post instead of using the retweet or reply function. This is done to avoid unwanted attention and evade text searchability \citep{burrell2019twitter}. Voldemorting refers to not mentioning words or names in a post, and instead inventing alternative names. This should likewise avoid a forced connection and unwanted attention \citep{nagel2018twitter}. 
Another interesting case of algorithmic leverage is the DoorDash \#DeclineNow movement, which will be the main subject of this work and discussed in depth in the next section.

\subsection{The Decline Now case}

DoorDash Inc.~is a US-based on-demand food delivery company, founded in 2013. As a gig economy company, it provides a digital platform for drivers, customers and merchants. After roughly doubling its market share since the Covid-19 pandemic, Doordash became the largest food delivery company in the US with a market share of 64\% \citep{secondmeasures2023competition}. In 2022 the DoorDash platform reports over 6 million drivers and roughly 32 million monthly active consumers \citep{sec2022doordash}. 
DoorDash drivers are paid per delivery, and usually deliver with their own car, bike or scooter, coming up for costs of car fuel and maintenance themselves. 

Very low-paying orders are a daily struggle for DoorDash drivers, with nearly a third of the orders paying less than \$0 after accounting for basic expenses~\citep{workingwanofreelunchreport}. In 2019 two DoorDash drivers, Dave Levy and Nikos Kanelopoulos, started the movement \#DeclineNow \citep{bloomberg2021doordash,vice2021doordash} which aims at tackling exactly these low-paying orders. Based on observations, they hypothesize that when a driver declines an order, the algorithm responds by reoffering the order to another driver with slightly increased pay by e.g. \$0.25. Thus, their strategy is to decline orders below a threshold of \$7 and to convince peer drivers to do the same. The movement is organized in Facebook groups and coined the hashtag \#DeclineNow. In order to better understand both the strategy and the reality of DoorDash drivers we spent a significant amount of time in these groups as part of our research.

Their hypothesis about the algorithm reminds of a reverse auction: If no driver wants to delivery the order at the given price, the price gets increased until a driver is willing to deliver it. However, by declining orders a driver cannot influence the overall pay level, they can only influence the pay for the subsequent driver on the same order. In this sense the \#DeclineNow strategy operates at a very local level, on a per-order basis. At the same time this means, that it must be a collective effort in order to be beneficial to all participants, as a single driver can not benefit by leaving orders to others. 

\subsection{Academic accounts of algorithmic levers} Despite the growing practical relevance, there is only a handful of academic papers that have formally studied algorithmic collective action strategies. \citet{vincent2021data} discuss how platform users can influence the effectiveness of AI powered technology by leveraging their control over data resources. Using simulations they demonstrate the effectiveness of withholding data through \emph{data strikes} \cite{vincent2019datastrike}, devalue data through \emph{data poisoning} \cite{tian2022comprehensive} and increase market competition through \emph{conscious data contributions} \cite{vincent2021conscious}. More empirically, \citet{laufer2022collective} formalize and quantify the strength of \emph{collective obfuscation} from a dataset of call logs in a real-world reporting hotline. \citet{creager2021online} demonstrate that algorithmic recourse can be improved through algorithmic collective action. \citet{hardt2023algorithmic} demonstrate that by combining strategic reporting of training data with collective strategizing at deployment time, collectives of small fractional size can gain significant control over a outcome of a platform's learning algorithm. \citet{baumann2024algorithmic} study algorithmic strategies in the context of music streaming platforms and illustrate unexplored potential to contest transformer-based recommendations through strategic interactions with the platform. 

Our work complements existing theoretical studies in the direction of collective strategizing against combinatorial algorithms, inspired by the real world case of the \#DeclineNow campaign. As pay data from gig workers and information about the platforms' algorithms are rarely available and often insufficient for a pure quantitative analysis, we propose a combinatorial model which we instantiate with empirical information.

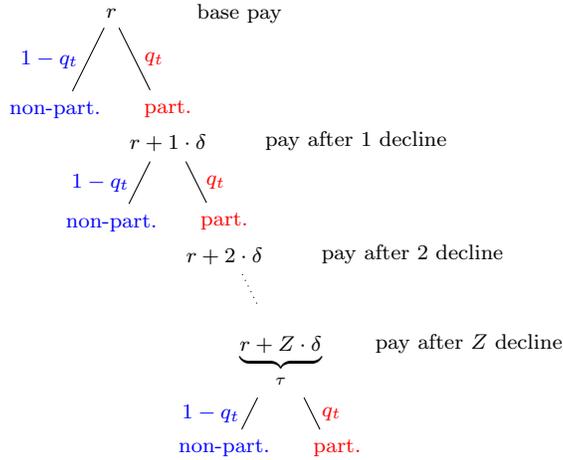
\begin{figure}[t]
\centering
\begin{tikzpicture} 
    \node (a1) {\scriptsize$r$}
        child {node[align=center, blue] {\scriptsize non-part.\\ } edge from parent node [left, blue] (N1) {\scriptsize$1-q_t$}}
        child {node (a2)[align=center, red] {\scriptsize part.\\ \scriptsize\color{black}$r+1\cdot \delta$} 
                child {node[align=center, blue] {\scriptsize non-part.\\ } edge from parent node [left, blue] {\scriptsize$1-q_t$}}
                child {node (a3)[align=center, red] {\scriptsize part.\\ \scriptsize\color{black}$r+2\cdot \delta$} 
                       child {node { } edge from parent [white] }
                       child {node (a4)[align=center]{\scriptsize \\ \scriptsize $\underbrace{r+ Z\cdot \delta}_{\tau}$}
                              child {node[align=center, blue] {\scriptsize non-part.\\ } edge from parent[solid] node[left, blue] {\scriptsize$1-q_t$}}
                              child {node[align=center, red] {\scriptsize part.\\ } edge from parent[solid] node[right, red] {\scriptsize$q_t$}}
                              edge from parent [dotted]}
                      edge from parent node [right, red] {\scriptsize$q_t$}}
                edge from parent node [right, red] {\scriptsize$q_t$} };
    \node[right of=a1, xshift=0.7cm]{\scriptsize base pay};
    \node[right of=a2, xshift=1.5cm, yshift =-0.2cm]{\scriptsize pay after 1 decline};
    \node[right of=a3, xshift=1.5cm, yshift =-0.2cm]{\scriptsize pay after 2 decline};
    \node[right of=a4, xshift=1.5cm, yshift =0.1cm]{\scriptsize pay after $Z$ decline};
\end{tikzpicture}
\caption{Assignment of orders in our model. In each level of the tree the order is assigned with probability $q_t$ to a participant and with probability $1-q_t$ to a non-participant. Every time a participant declines the order, the pay increases by an extra $\delta$. An order can be declined maximally $Z= \frac{\tau - r}{\delta} $ times, before the threshold $\tau$ is reached and it is accepted by both groups. The probability $q_t$ can vary in different time steps $t$. The figure illustrates a time step with $I_t=1$. }
\label{fig:tree}
\end{figure}

\section{Model description}
\label{sec:model}

We assume there is a fixed population of $N$ workers on the platform. Each worker can be either busy with delivering an order or idle. In each time-step $t$ there is one incoming order. The platform assigns each incoming order randomly to one of the idle workers. Each order has an initial base pay of $r$.  If the order is declined the pay is raised by an extra $\delta$ and the order is reassigned randomly among the idle workers – including the worker who just declined the order. This random reassignment of the order happens until the order is accepted. If a worker accepts an order they are busy for a period of $b$ time-steps – which we assume to be constant for every order. We use $n$ to denote the number of orders per hour.~\looseness-1

\paragraph{Collective action strategy of declining.} We assume participating workers agree on a threshold $\tau\geq r$ and follow the strategy $h_{\tau}$ to decline an order unless the pay meets the threshold: 
\begin{align*}
    h_{\tau}(\text{pay}) = 
    \begin{cases}
        \text{decline, } & \text{if pay} < \tau\\
        \text{accept}  & \text{otherwise.}
    \end{cases}
\end{align*}
Non-participating workers accept an order at any pay. We use $\alpha\in[0,1]$ to denote the fraction of workers participating in collective action and implementing the strategy.

\subsection{Order assignment}
Each order is randomly assigned to an idle worker, either a participant  or a non-participant. If the order is declined the pay is increased by $\delta$ and the order is reassigned. If the order is accepted the next order is processed.
Figure~\ref{fig:tree} illustrates how the pay evolves, where $q_t$ denotes the probability that an order is assigned to one of the participating workers at time step $t$ and corresponds to the fraction of participants among the idle workers competing for the order.

\paragraph{Zero-supply time steps.} An order can only get delivered if at least one of the workers is idle at that time step. When all workers are busy, the order will not get delivered. We use the following indicator variable to describe the two cases: 
\begin{align}
    I_t := 
    \begin{cases}
        0 & \text{no idle worker present}\\
        1 & \text{idle worker(s) present} \:.
    \end{cases}
\end{align}

\paragraph{Who gets the order?} 
We are interested in the probability that a participant of the collective \textit{accepts} an incoming order for the case where $I_t=1$. Since all participants of strategy $h_{\tau}$ accept orders only if the pay matches their threshold $\tau$, an order has to be assigned to and declined by participants $Z= \frac{\tau - r}{\delta}$ times – we assume $\tau - r$ to be a multiple of $\delta$ such that $Z\in \mathbb{N}$.
Hence, the overall probability for participants to accept the order at time-step $t$ under $I_t=1$ is given by 
\begin{equation}\beta_t := q_t^{Z+1} \quad \text{with }\quad Z = \frac{\tau - r}{\delta}  \:.
\label{eq:Pt}
\end{equation}

Analogously for non-participants the overall probability to accept an order at time-step $t$ is $1-\beta_t$. This corresponds to the cumulative probability of ending on one of the blue leaves in the tree of Figure~\ref{fig:tree}.
The fraction of participants among idle workers, $q_t$, is defined recursively by accounting for the workers who have been assigned orders at the previous time step and the formerly busy driver who is joining the pool of idle drivers after $b$ time-steps.

\paragraph{The spillover effect.} Participants only accept the order if the pay reaches $\tau$. Non-participants accept an order at any pay. However, the pay for non-participants can increase due to participants having declined the same order previously. We refer to the expected pay increase for a non-participating worker as the \emph{spillover effect}. The spillover effect at time-step $t$ is denoted by $R_t$. In Figure~\ref{fig:tree} this corresponds to the expected increase over the base pay $r$ across all branches where a non-participant accepts the order:
\begin{equation}
\label{eq:Rt}
\begin{split}
    R_t &:= \mathds{E} \big[\text{pay increase} \mid \text{non-participant accepts order at time $t$}\big] \\
    &:= \sum_{z=0}^{Z} z\delta \: \frac{q_t^z (1-q_t)}{1-q_t^{Z+1}} = \frac{\delta q_t(Zq_t^{Z+1} - (Z+1)q_t^Z +1)}{(1-q_t)(1-q_t^{Z+1})}  \:.
\end{split}
\end{equation}
The expectation is over the randomness in the assignment, the derivation can be found in Appendix \ref{sec:proof_eq_Rt}.
Note that for all orders that end up being assigned to non-participants we must have $q_t < 1$. Thus, the spill-over effect is always positive and $R_t\in(0,\tau-r)$. 

\subsection{Worker utility}

We define the utility for a worker as the expected hourly income. This corresponds to the expected cumulative pay for all the orders they accept per hour, minus the overheads for delivering. To specify overheads we use $m$ to denote the average miles per order and $c$ to denote the average costs per mile, which yields an overhead of $cm$ per order. To specify the revenue we consider participants and non-participants separately.

\paragraph{Participating workers.} By definition of our model, participating workers only accept orders if the pay exceeds a threshold $\tau$. Thus, they get a pay of $\tau$ for every accepted order. Using notation established in the previous subsection this yields a utility per participant of
\begin{align} \label{eq:uD}
u^D(\alpha, \tau) 
&:= \frac 1 {\alpha N} \; (\tau-cm) {n \mathds{E}_t \left[\beta_t I_t \right]}\:.
\end{align}
We use the superscript $D$ for participants to indicate that they decline orders.

\paragraph{Non-participating workers.} Non-participating workers get the base pay per order, plus the benefit from spill-over effects. Again, building on notation from the previous subsection, this yields an hourly income for non-participants of
\begin{align}
\label{eq:uA}
u^A(\alpha,\tau) 
&:= \frac 1 {(1-\alpha) N} \; (r-cm) {n\left(1-\mathds{E}_t [\beta_t I_t ]\right)}+ \frac n {(1-\alpha) N} \;{R} 
\end{align}
where  $R$ denotes the expected spill-over effect defined as
\begin{align}
R:={\mathds{E}_t \left[(1-\beta_t)R_t I_t \right]}
\end{align}
Note that $R_t$ and $\beta_t$ also indirectly depend on $\alpha, \tau$. While $\beta_t$ is monotonically increasing in $\alpha$, and monotonically decreasing in $\tau$, the opposite is true for $R_t$. The superscript $A$ indicates that non-participants accept order. Note that $u^D$ and $u^{A}$ are computed per worker.

\paragraph{Average worker utility.} 
Lastly, we are interested in the average utility across both participating and non-participating workers, which we denote as
\begin{align}
\label{eq:u_average}
u(\alpha, \tau):= \alpha \cdot u^D(\alpha,\tau) + (1-\alpha) \cdot u^{A}(\alpha,\tau) \:.
\end{align}
It serves as a proxy for the average gain for the workforce. 

\subsection{Market conditions}
\label{subsec:marketconditions}

We identify the supply condition as a key property of our model. Oversupply of labor relative to demand is predominantly present throughout the gig economy and plays a crucial role in the collective power of workers [\citealp{graham2019global}, \citealp{woodcock2019gig}]. Within our model, we can characterize the two regimes of over- and undersupply, as well as the degree of oversupply by the following definitions:

\begin{definition}[Supply-Conditions]
\label{def:supplycond}
    Define an undersupply of labor if $N \leq b$. 
    Conversely, define an oversupply of labor whenever $N > b$.
\end{definition}

\begin{definition} [Degree of Oversupply]
\label{def:degreeofoversupply}
    The degree of oversupply is defined  as the factor of supply relative to demand: 
    \[\deg := \frac{N}{b}.\] 
    It holds that
    $\deg = 1$ represents a matching supply, $\deg<1$ corresponds to undersupply and $\deg>1$ corresponds to oversupply.
\end{definition}

To interpret this, recall that $N$ is the total number of workers, representing the supply of labor. In turn, $b$ is the busy time and characterizes indirectly the demand by determining how many workers are needed to serve a continuous stream of orders. In this sense $\deg$ measures the factor of oversupply of labor. Our key results will focus on the phase transition in our model implied by these definitions. 

\section{Simulation and parameter instantiation}
\label{sec:paraminstantiation}

Our model is defined by a range of exogenous variables. In the following we will discuss plausible instantiations for each variable that we will use for our simulations. To this end, we rely on news articles, public artifacts, and a report from \cite{workingwanofreelunchreport}. The latter collected over 200 data points from DoorDash drivers and contains data on pay, delivery duration and tips, as well as other external sources. These are rare exceptions of data on worker pay and market conditions being available. Together with our model we can use this data to effectively study the strategy. We chose the following instantiations:
\begin{itemize}[noitemsep]
    \item \textbf{Acceptance threshold} ($\tau=7$): This corresponds to the threshold of the \#DeclineNow movement which was set to \$7 [\citealp{vice2021doordash}, \citealp{bloomberg2021doordash}].
    \item \textbf{Base pay for an order ($r=4$):} The minimum pay for one order is at \$2, see  \citep{doordashhelp}, which according to \citet{workingwanofreelunchreport} makes up 8\% of the orders. As the model considers only low-paying orders below the threshold $\tau$, we chose $r$ to be in between, such that $r > cm$. 
    \item \textbf{Average cost per mile} ($c=0.6$): The Internal Revenue Service in the US proposes roughly 60ct/mile for car fuel and maintenance \citep{irsmileage}.
    \item \textbf{Average miles per order} ($m=6$): As the model considers only low-paying orders we assume it to be a little less than the 6.8 miles per order calculated by \citet{workingwanofreelunchreport}.
    \item \textbf{Average number of orders per hour} ($n=60$): This is an  estimate for the number of orders per hour within an algorithmically relevant radius (i.e. within a radius such that the algorithm would still assign it to the same driver). We could not find documented information on this.
    \item \textbf{Average number of time-steps delivering an order} ($b=30$): \citet{workingwanofreelunchreport} report an average of 30 minutes per order on DoorDash. 
    \item \textbf{Pay increase after decline} ($\delta=0.25$): This means a pay increase of \$0.25 each time an order is declined as it is often reported in forums such as Reddit. 
\end{itemize}

We will use these values for our simulations of the dynamics across time. When simulating the models, we first perform a warm-up phase of $5  b$ time-steps until the model reaches equilibrium and then inspect 10,000 time-steps. In our experiments we vary the group size $\alpha$ in $[0,1]$ and the total number of drivers $N$ between 30 and 150, implying that the number of drivers that can influence each other’s pay by declining an order is in the range of less than or a few hundred. By varying $N$ and holding $b$ fixed we vary the degree of oversupply.

\section{Key results}
\label{sec:results}

Our main results investigate under which conditions the strategy is beneficial – either averaged over the aggregate work force, or for the group of participants – and relate success to the size of the collective and the degree of oversupply. 

\paragraph{Reference point and intuition.}
As a reference point we consider the scenario without any strategy where orders are randomly assigned to one of the idle workers.
We denote the resulting utility the \emph{base utility}, $u^{\text{base}}$.  

The base utility is constant in the case of undersupply as workers are busy at every time step, and inversely proportional to the degree of oversupply once $N>b$ and workers compete for orders. In our model the base utility corresponds to the setting where the threshold $\tau$ is set to $r$, or alternatively, the case where $\alpha=0$. 

\begin{figure}[t]
  \centering
  \includegraphics[width=1.0\textwidth]{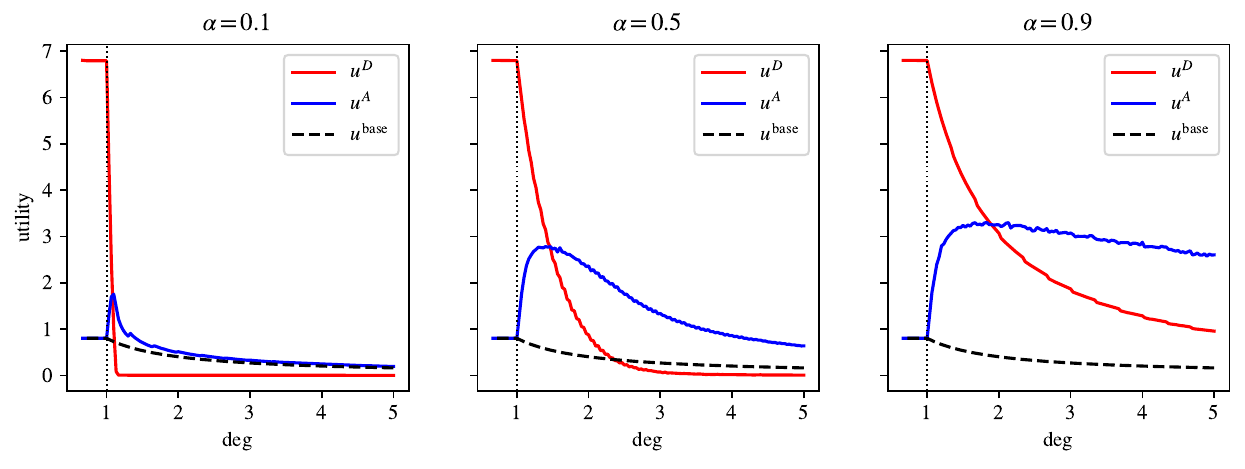}
  \caption{Utility and tradeoffs. Base utility (dashed black), utility for participants (red) and non-participants (blue) for different levels of participators, $\alpha$. deg $< 1$ represents an undersupply and the dotted vertical line marks the transition from under- to oversupply. }  \label{fig:utilities}
\end{figure}

We simulate $u^{\text{base}}$, $u^{A}$ and $u^D$ empirically in Figure~\ref{fig:utilities} as a function of the degree of oversupply for different levels of $\alpha$. Figure~\ref{fig:utilities} shows that in the regime of undersupply (deg$<1$) the utility for participating workers is larger than the utility for non-participating workers, for all three levels of $\alpha$. In the case of oversupply, this relationship flips, being worse for larger degrees of oversupply and smaller levels of participating workers, $\alpha$. The figure exemplifies that there is a regime where participating is self-incentivized and a regime where this is not the case. Since such an incentive structure has important implications for the organization of a collective action we characterize the individual utilities and trade-offs in the following.

\subsection{The gain of collective action}

First, we inspect the benefit for the overall  workforce from engaging in collective action: How does the utility for an average worker change when a collective decides to engage in the strategy? We define the gain of collective action as follows: 

\begin{definition}[Gain of Collective Action] 
    The gain of collective action is defined as the difference between the average utility under the strategy and the utility without any strategy: 
    \[G(\alpha, \tau) := u(\alpha,\tau) - u^{\text{base}}\]
    where $u(\alpha, \tau)$ denotes the average utility across all workers, defined in \eqref{eq:u_average}.
\end{definition}

\begin{theorem}[Overall Gain of Collective Action] \label{th:gainpositive}
    For any degree of oversupply, and for all $\alpha > 0, \tau > r$ the gain of collective action is positive:
\[G(\alpha, \tau) > 0. \] 
\end{theorem}

This implies that $G$ is always positive, regardless the group size $\alpha$ or the supply conditions. Thus, on average the workers are better off with collective action than without. As participating workers decline, payments go up and someone benefits.

\begin{figure}
  \centering
  \includegraphics[width=0.55\textwidth]{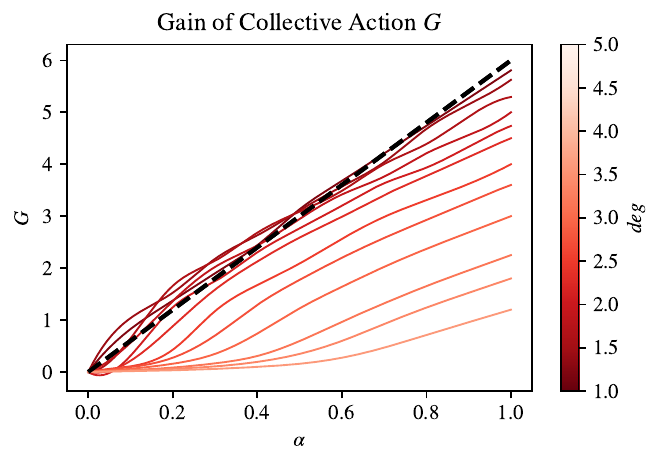}
  \caption{Gain of collective action for varying supply level $\deg$ and collective size $\alpha$. The dashed black line indicated $\deg\leq 1$ and the colored lines the different supply levels. We fix $b=30$ and vary $N \in [30, 150]$.}  \label{fig:gain} 
\end{figure}

\paragraph{Dependence on group size and supply level.} To gain further insights, we support our results with simulations of the model. Figure \ref{fig:gain} illustrates the gain of collective action for different group sizes $\alpha$ and degrees of oversupply, indicated by the color of the lines. The dashed line indicates $\deg\leq 1$.  
We can observe three interesting aspects: 
\begin{enumerate}[label=(\roman*), itemsep=0pt]
\item 
In the case of undersupply, the gain increases linearly in $\alpha$.
    \item For large $\alpha$ the gain of collective action decreases with the degree of oversupply. 
    \item For small values of $\alpha$, a small degree of oversupply can be beneficial over the regime of undersupply (everything above the black line in Figure \ref{fig:gain}).
\end{enumerate}
The first relationship can be read of the proof of Theorem \ref{th:gainpositive} and is depicted by the black dashed line in Figure \ref{fig:gain}. The second effect makes intuitively sense since the overall gain $G$ is an average measure per worker it shrinks with $N$. The third aspect is due to the fact that it needs a little bit of oversupply for the price increase due to declining to spill over to the non-participating group. This effect increases the overall gain. With increasing $N$ this spillover effect $R$ eventually cancels with the effect of reducing the average workload. 

\begin{figure}[t]
  \centering
  \includegraphics[width=0.55\textwidth]{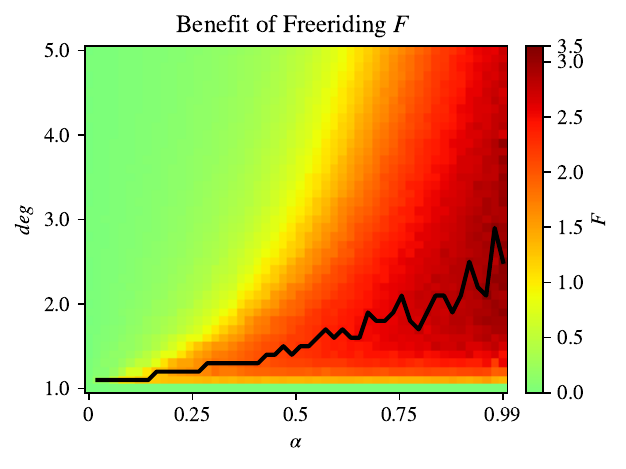}
  \caption{Benefit of freeriding for varying supply level $\deg$ and collective size $\alpha$. We fix $b$ and vary $N \in [30, 150]$. The black line shows the maximum $F$ for each $\alpha$.} 
  \label{fig:freeriding}
\end{figure}

\subsection{Benefit of freeriding}
\label{sec:freeriding}

Next, we are interested in how non-participants benefit from freeriding. Freeriding generally refers to benefiting from a public good without contributing to it. In our model, non-participants benefit from the increased price per order, without declining orders themselves. We formally define the benefit of freeriding by comparing the utility of non-participants to the utility when nobody declines: 

\begin{definition}[Benefit of Freeriding]
    The benefit of freeriding is defined as the difference between the utility for non-participants and the base utility: 
    \[F(\alpha, \tau) := u^A(\alpha,  \tau) - u^{\text{base}}.\]
\end{definition}

\begin{theorem}\label{th:freeriding}
    In the case of undersupply, i.e. $N\leq b$, the benefit of freeriding is 0: 
    \[F(\alpha, \tau) = 0.\] 
\end{theorem}

This theorem shows that in the case of undersupply there is no benefit to freeriding, as there is no spillover effect.
Beyond undersupply, simulations depicted in Figure \ref{fig:freeriding} show that freeriders gain the most in case of large groups of participants and low degrees of oversupply which is the scenario where the strategy is most effective.

\subsection{Benefit of participation}
\label{sec:benefitofparticipation}

Lastly, we are interested under what conditions it is beneficial to participate in the strategy. From a game theoretic viewpoint, this amounts to asking when the expected utility for participating is larger than for not participating. We use the following definition: 

\begin{definition}[Benefit of Participation]
    The benefit of participating in the strategy is defined as the difference between the utility for participants and non-participants: 
    \[B(\alpha, \tau) := u^D(\alpha,  \tau) - u^{A}(\alpha, \tau).\] 
\end{definition}

A first result shows that for the case of undersupply, it is always beneficial to participate: 

\begin{theorem}
\label{th:undersupplybenefitpositive}
In the case of undersupply ($N\leq b$) the benefit of participation is positive and given by:
\[B(\alpha, \tau) = \frac{n}{b} (\tau - r) > 0 .\] 
\end{theorem}

Hence, in the case of undersupply non-participants are worse off than participants. Thus, it is individually rational to participate. 
This no longer holds for the case of oversupply where non-participants can benefit from spill-over effects. We ask under which conditions the benefit of participation is still positive:

\begin{figure}
  \centering
  \includegraphics[width=0.55\textwidth]{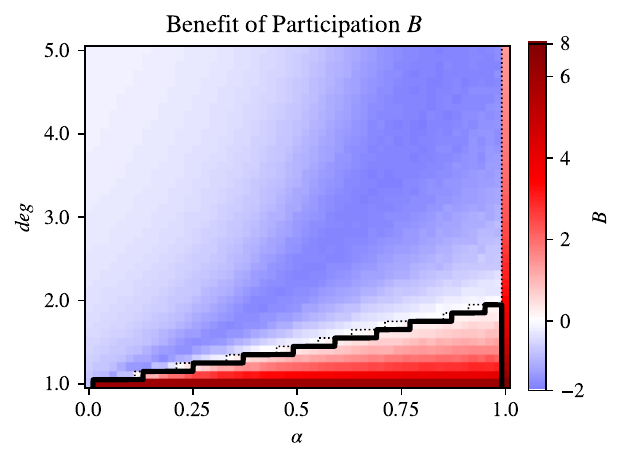}
  \caption{Benefit of participation for varying supply level $\deg$ and collective size $\alpha$. We fix $b=30$ and vary $N \in [30, 150]$. The black line shows the boundary of the condition from Theorem \ref{th:oversupplygroupsizethreshold} under which $B>0$. The dashed black line shows the ground-truth boundary obtained from the simulation.}
  \label{fig:benefit}
\end{figure}

\begin{theorem}[Threshold of Participation]
\label{th:oversupplygroupsizethreshold}
Assume that $r>cm$. Then, in case of oversupply, the following condition on degree of oversupply $\deg$, group size $\alpha<1$ and threshold $\tau$ has to be fulfilled such that the benefit of participation is positive: 
\[ \deg < \frac{1}{1-\alpha} \Big ( 1- \frac{\alpha (r+R-cm)}{(1-\alpha)(\tau - cm) + \alpha (r-cm)} \Big ) \Rightarrow B(\alpha, \tau) > 0 .\]  
\end{theorem}

The theorem shows that it is sufficient that the degree of oversupply is small and the group size $\alpha$ is large, then the benefit of participation is positive in the regime of oversupply. The assumption of $r>cm$ means that the base pay covers for the costs. 

The theorem captures the point in Figure \ref{fig:utilities} where the participant utility crosses the non-participant utility. As Figure \ref{fig:utilities} suggest, the smaller $\alpha$ the less oversupply is permissible to be in the beneficial regime for participants. This is an economically significant regime, as it means that participation is self-incentivized.

\paragraph{Dependence on group size and supply level.} To augment the result from the theorem we simulate our model to depict the benefit of participation for different values of group size $\alpha$ and degrees of oversupply, see Figure \ref{fig:benefit}. On the one hand, this reveals that the boundary implied by the condition from Theorem \ref{th:oversupplygroupsizethreshold} (thick black line) is relatively tight compared to the ground-truth obtained from the simulation (dashed black line). On the other hand, we see yet again that both a large fractional size $\alpha$ and a small degree of oversupply is crucial to make the strategy remunerative for a participant.

\begin{figure}[t!]
\centering
\includegraphics[width=.29\textwidth]{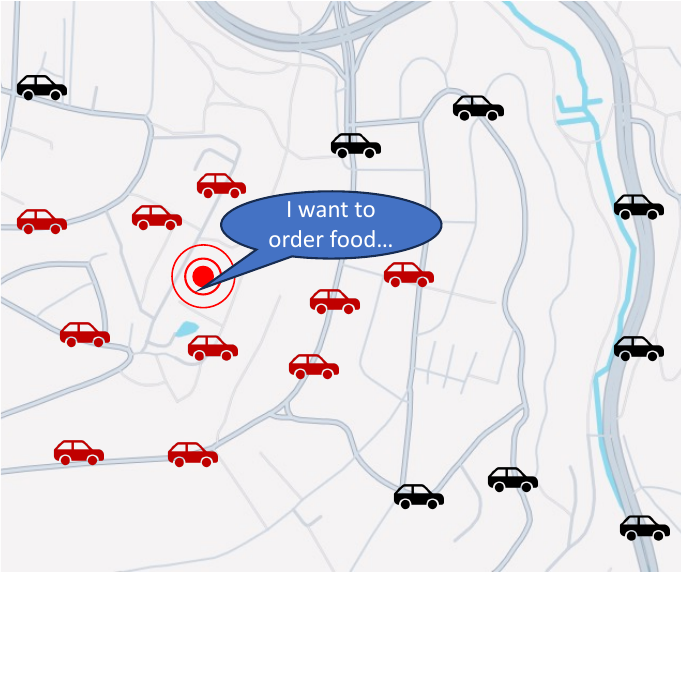}\hfill
\includegraphics[width=.29\textwidth]{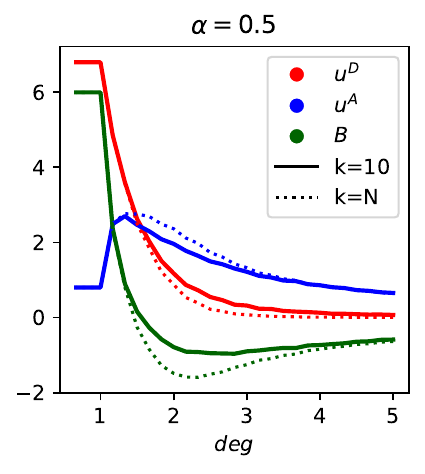}\hfill
\includegraphics[width=.39\textwidth]{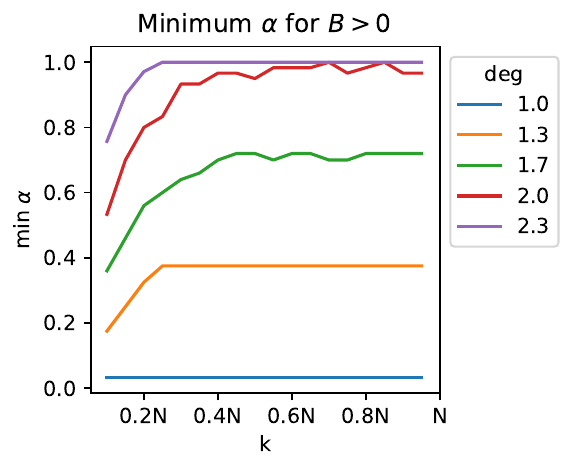}
\caption{Simulating locality of order  assignment in our model. \textbf{(left)} Visualization how an order affects the $k$ nearest drivers, for $k=10$. 
\textbf{(middle)} For $k=10$, utility for participants (red) and non-participants (blue) are shown for different degrees of oversupply, with a fixed $b=30$. 
The resulting benefit of participation is shown in green. Utilities and benefit of participation for global assignment ($k=N$) is shown in dotted lines. \textbf{(right)} Smallest $\alpha$ such that the benefit of participation is positive, for varying $k$ on the x-axis, and different degrees of oversupply as shown in the legend, with fixed $b=30$.}
\label{fig:robustness}
\end{figure}

\subsection{Robustness to model misspecifications}

Ordering food and assigning the order to nearby drivers happens presumably at a very local scale throughout a city. To test for the robustness of our findings with regard to the locality of order assignment, we extend our model with a parameter $k$. It indicates how many workers around the order dispatch location are being considered when the order is assigned, see Figure \ref{fig:robustness} (left). That is, for each order we randomly sample $k$ workers from the pool of idle workers. In particular this means, that in each order assignment the number of participants and non-participants may vary. We recover the base model for $k=N$. 

Our robustness simulations corroborate our previous results. Figure \ref{fig:robustness} (middle) shows that for $k=10$ the participant utility (red) is slightly increased, the non-participant utility (blue) is slightly decreased and the resulting benefit of participation (green) is higher as compared to $k=N$ (dotted). The rightmost panel of Figure \ref{fig:robustness} shows the smallest $\alpha$ for which the benefit of participation becomes positive, for varying $k$ and for different degrees of oversupply. It demonstrates that with decreasing $k$ the minimum $\alpha$ gets smaller – which in turn means that the regime of positive benefit is reached for lower group sizes. 

The ablation indicates that overall trends are preserved under the inspected deviations from the model, having little effect on the qualitative take-aways from our results. 
Further, our model tends to offer a conservative account of participant utility, likely preserving the validity of our sufficient conditions on the threshold for participation.

\begin{figure}[t!]
\includegraphics[width=\textwidth]{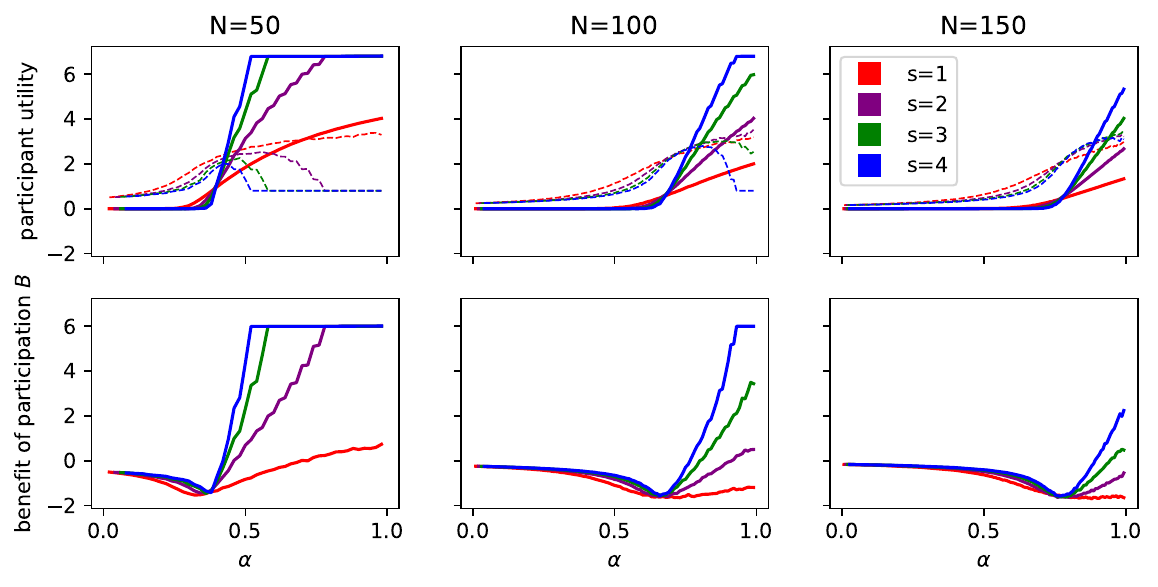}
\centering
\caption{Working in shifts. \textbf{(top)} Utility under shift the strategy for varying fractions of participating workers $\alpha$ and total number of workers $N \in \{50, 100, 150\}$. Shift utility for participants displayed with solid line, utility for non-participants dotted line. \textbf{(bottom)} Benefit of participation $B$. }
\label{fig:shiftstrategy_N}
\end{figure}

\section{Working in shifts}
\label{sec:workinginshifts}

We have established, that collective action is particularly effective in the regime where the degree of oversupply is low. A concrete recommendation that emerges from our results is that workers should target regimes of low degree of oversupply, and ideally enter the regime of undersupply. In reality however, large degrees of oversupply are often the predominant scenario, as documented in several studies. \cite{graham2019global} report roughly 90\% oversupply for the freelance platform Upwork. \cite{suhr2019two} report 200-650 ridehail drivers for less than 75 jobs in a large Asian city. In this section we explore what practical action the collective can take to get into the more beneficial regime of supply, where collective action is more effective. 

A way to implement this is to organize as a collective and work in different shifts, instead of working all at the same time. By doing so, the collective can influence the supply of labor. Within our model the strategy can be implemented as follows: Assume a collective of size $\alpha$. Say each worker is willing to work for a fixed amount of hours per day. Let $s$ be the number of shifts the collective wants to implement, the collective could work in one ``shift", $s=1$, or to work distributed over the day in up to $s=4$ shifts, each worker being active for a total of $6$ hours. 
Working in multiple shifts means less participants per shift and thus a smaller degree of oversupply. By working in $s$ shifts  the total number of workers is reduced from $N$  to $\widetilde{N} = N- (\alpha - \frac \alpha s)N$, directly translating into an effective degree of oversupply of $\tilde {\deg}=\frac{\tilde N} b$. The degree of oversupply decreases with more shifts $s$ and larger $\alpha$ (see additional Figure in Appendix \ref{sec:additionalfigureshifts}).  
However, this comes with a trade-off. At every point in time there are fewer active workers who decline, relative to accepting. Thus,  the effective collective size is reduced to $\widetilde{\alpha} = \frac{\alpha}{s}$. The key question is, what is more beneficial – having more ``decline-power" or less degree of oversupply? 

\paragraph{Simulation.}
To answer this question, we simulate the shift strategy in our model. We use  $N\in \{50, 100, 150\}$ and $s = 1, ..., 4$.  The top row in Figure \ref{fig:shiftstrategy_N} shows how the number of shifts impacts the utility structure in the model. The leftmost panel shows that above a certain $\alpha$ having as much shifts as possible yields the most utility for a participant. The middle and rightmost panel of Figure \ref{fig:shiftstrategy_N} show that for larger $N$ this $\alpha$-threshold has to be larger, as degrees of oversupply are larger thus it is harder to lower them significantly. The bottom row in Figure \ref{fig:shiftstrategy_N} shows the resulting benefit of participation, $B$.

Essentially, the shift strategy points out the trade-off between a low degree of oversupply and having a large collective to increase the probability of orders being declined. Our results indicate that the additional layer of coordination can be a very effective strategy for larger collectives.

\section{Discussion}
\label{sec:discussion}

Our model outlines the benefit of collective action at any group size and identifies the level of supply as an important factor driving the success of collective action. These findings are further be supported with empirical evidence.

In fact, there is empirical evidence that the strategy of \#DeclineNow works: \citet{vice2021doordash} mentions an experiment carried out by drivers organized in a Facebook group. In one experiment they accepted every order while in another experiment they selectively declined low-paying orders. Results were that drivers who accepted everything earned \$0.68 per mile with a gross hourly wage of \$12.41. Drivers who selectively declined earned \$1.26 per mile with a gross hourly wage of \$21.15.

Further, high degrees of labor oversupply are a real issue in the gig economy and are held responsible – among other factors – for low wages and little bargaining power~\citep{graham2019global}.  
Nevertheless, presumably as long as workers are classified as independent contractors the lever of declining to raise pay is real and valuable – and not necessarily restricted to DoorDash. The question is how to apply and organize it, especially in the light of ever more personalized wages in the gig economy \citep{dubal2023algorithmic, teachout2023algorithmically}. Out of curiosity, we asked in a \#DeclineNow Facebook group whether this strategy is also applicable to other gig platforms. Answers pointed in the direction that ``all the gig apps do add extra money if the order isn’t taken”. One commentator nevertheless highlighted that ``these companies intentionally keep rotating in new people all the time that don’t know any better so they accept these offers. That is literally the business model”. James Farrar, a gig worker activist and the founder of Worker Info Exchange, twittered in 2023 that ``withholding labour at Deliveroo works! Yesterday, having refused a delivery job for 90 minutes, Deliveroo’s algorithm `Frank’ steadily increased the pay offer by 52\% from £6.04 gross to £9.20”\footnote{\url{https://x.com/jamesfarrar/status/1739653646437032286?s=20} accessed on 09-11-2024}.

We explored the idea of working in shifts as a suggestion to reduce oversupply and demonstrate its effectiveness under our model. However, implementing such a measure would require more coordination than the declining strategy alone: All participating workers have to coordinate when to work, whereas for the declining strategy participants can act individually. Thus, tools and apps for organizing participants~\citep[c.f.,][]{kimberly24,Toxtli2023}, will certainly play an important role in improving work conditions for gig workers.

Developing tools for drivers in the context of ride hailing to coordinate their work-times, and testing them in practical scenarios would be an interesting direction for future work. In particular, the app should facilitate coordination, and build trust among participants to counter the financial incentives of platforms not to engage in such strategies. We envision modeling insights like ours could complement the scarce data about pay and platform algorithms to guide and facilitate the development and dissemination of new strategies.

\section*{Acknowledgements}

The authors would like to thank Ana-Andreea Stoica and Jiduan Wu for valuable feedback on the manuscript. Celestine Mendler-Dünner acknowledges financial support of the Hector Foundation.

\bibliographystyle{plainnat}
\bibliography{references}

\newpage
\appendix

\section{Auxiliary results}

\begin{lemma}
\label{lemma:undersupply_nb}
We use 
$\gamma_D := \frac {n^D}{n^A+n^D}$ to denote the participant fraction of delivered orders. It holds that
\begin{itemize}
\item In the case of \textbf{undersupply:} \quad $\gamma_D=\alpha$
\item In the case of \textbf{oversupply:} \;\;\;\quad 
$\gamma_D \in\left[ \textnormal{max } \Big(0, 1-(1-\alpha) \cdot \deg\Big), \alpha\right]$
\end{itemize}
\end{lemma}

The intuition is that in the case of undersupply every worker gets an order as soon as they become idle, thus $\gamma_D$ reflects the fractional group size of participants. In the case of oversupply, participants get orders either by chance (i.e. being offered orders at threshold pay) or by being a large enough group such that they can determine the pay when the few non-participants are busy, i.e. competition-free time steps. Such time-steps exist as long as $(1-\alpha)N<b$, or in other words $\deg\leq \frac 1 {1-\alpha}$. The lower bound characterizes exactly these competition-free time steps.

\begin{lemma}
\label{lemma:R_undersupply}
We denote the expected pay increase for non-participants, i.e. the spillover effect $R := \mathds E_t [(1-\beta_t)R_t I_t]$. It holds that 
\begin{itemize}
\item In the case of \textbf{undersupply:} \quad $R=0$.
\item In the case of \textbf{oversupply:} \;\;\;\quad 
$R>0, \;\; \forall \: 0 < \alpha < 1.$
\end{itemize}
\end{lemma}

This result follows from the observation that in the case of undersupply there is no competition around orders, and thus non-participating drivers always receive the base pay $r$. This is different to the case of oversupply where non-participants potentially benefit from competing participants declining orders. 

\begin{lemma}
\label{lemma:baseU}
The base utility without any strategy is characterized as 
\begin{equation}\label{eq:u0}
    u^{\text{base}} :=  \frac 1 {\min(\deg,1)}\frac{n}{b } (r-cm) \:.
\end{equation}
\end{lemma}

\section{Proofs}

\subsection{Proof of Lemma~\ref{lemma:undersupply_nb}}

For the case of undersupply, we want to prove $\gamma_D = \alpha$. As every worker gets an order every $b$ time steps, a worker gets $\frac{n}{b}$ orders per hour. To account for group sizes, the participants get per hour $n^D = \alpha N \frac{n}{b}$ and non-participants $n^{A} = (1-\alpha)N \frac{n}{b}$. Then it holds: 
\[ \gamma_D = \frac{n^D}{n^D+n^{A}} = \frac{\alpha N \frac{n}{b}}{\alpha N \frac{n}{b} + (1-\alpha)N \frac{n}{b}} = \alpha \]

For the case of oversupply, the upper bound $\gamma_D \leq \alpha $ follows the same logic given by the simple fact that the participant fraction of delivered orders cannot be larger than the group itself.
It remains to prove the lower bound $\gamma_D \geq \text{max}(0, 1-(1-\alpha) \cdot \deg)$. In the case of oversupply, all orders get delivered and it holds $\gamma_D = \frac{n^D}{n}$. Consider a period of length $b$. Assume the worst case, namely that all non-participants get an order during $b$ and participants get the rest (if any left). If the group of participants $(1-\alpha)N$ is larger than $b$, participants do not get any order. If the group of participants $(1-\alpha)N$ is smaller than $b$, then there are $b-(1-\alpha)N$ competition free time steps. Divided by $b$ to compute per time step, and multiplied by $n$ this is a lower bound to $n^D$, which can be rearranged to the stated bound: 
\[n^D \geq n\cdot \text{max}\left(0, \frac{b-(1-\alpha)N}{b} \right) \Leftrightarrow \gamma_D \geq \text{max}(0, 1-(1-\alpha) \cdot \deg)\]

\subsection{Proof of Lemma~\ref{lemma:R_undersupply}}
\label{sec:proof_R_oversupply}

For the case of undersupply, we want to show that $R=0$. 
In case of undersupply of labor ($N \leq b$) each worker is getting the next order directly after having completed one, there is no competition around orders. This means, in each time-step there is at most one worker present, who is then getting the order. For all time-steps $t$ in which this worker is a non-participant the probability that a participant gets the order is 0: $q_t = 0, I_t=1.$ Recall that $R_t$ is only defined for $q_t < 1$, so the case where only a participant is present can be neglected here. For the case that no worker is present: $I_t=0$. \\
Thus: \[R = \mathds{E}_t [(1-\beta_t)R_t I_t] = \mathds{E}_{t} \left [ (1-q_t^{(Z+1)}) \cdot \frac{\delta q_t(Zq_t^{(Z+1)} - (Z+1)q_t^Z +1)}{(1-q_t)(1-q_t^{(Z+1)})} \cdot I_t \right] = 0\]

For the case of oversupply, we want to prove that $R>0$ for $0<\alpha <1$. 

We first introduce the notation $D_t$ for the number of idle participants in time-step $t$, and $A_t$ for the number of idle non-participants. For convenience we define the quantity of idle drivers in the case of oversupply as $N_{idle} := D_t + A_t$. 

To prove the lemma, we first show, that for $N > b$ and $\forall 0<M<N$: $\exists q_t \text{ such that } 0 < q_t < 1$.
Because it holds that $q_t$: $q_t = \frac{D_t}{D_t+A_t} = \frac{D_t}{N_{idle}}$ we can alternatively prove that $\exists D_t \text{ such that } 0 < D_t < N_{idle}$. 

Consider a time-period of length $b$ and the number of idle participants in each step $D_{t-b}, ..., D_t = 0$. Proof by contradiction:
\begin{itemize}
    \item Assume all $D_{t-b}, ..., D_t = 0$. We know that $M \geq 1$, w.l.o.g. assume that $M = 1$, i.e. there is one participant. Even if this participant were fully busy, there would be a time-step $t' \in \{t-b, ..., t\}$ where he would return from his previous order and would be available to catch up a new one. Thus $D_{t'} = 1$ which contradicts the assumption. Therefore for every time-period of length $b$ $\exists D_t > 0$, thus in expectation over time the probability that this $D_t$ occurs is non-zero, in particular $P(q_t) \geq \frac{1}{b}$.
    \item Assume all $D_{t-b}, ..., D_t = N_{idle}$. Recall that $N-M$ describes the group size of non-participants. We know that since $M < N$ there is at least one non-participant, i.e. $(N-M) \geq 1$ and assume w.l.o.g. that $(N-M) = 1$. The argument is the analogous: Even if this non-participant was fully busy, there would be a time-step $t' \in \{t-b, ..., t\}$ where he would return from his previous order and would be available to catch up a new one. Thus $A_{t'} = 1$ and $N_{idle} = D_{t'} + A_{t'} = D_{t'} +1 > D_{t'}$. Therefore for every time-period of length $b$ $\exists D_t < N_idle$, thus in expectation over time the probability that this $D_t$ occurs is non-zero, in particular $P(q_t) \geq \frac{1}{b}$.
\end{itemize}

\noindent
Therefore for every period of length $b$ it holds that $\exists D_t \text{ such that } 0 < D_t < N_{idle}$. 

\vspace{\baselineskip}
\noindent
Secondly, with that we can prove that $R > 0$. As we are in the case of oversupply we can we can ignore the indicator variable $I_t$ as it is always 1. We first do the following transformation:

\begin{align*}
    R &= \mathds{E}_t [(1-\beta_t)R_t I_t] \\
    &= \mathds{E}_{t} \Big [ \big(1-q_t^{(Z+1)}\big) \cdot  \frac{\delta q_t(Zq_t^{(Z+1)} - (Z+1)q_t^Z +1)}{(1-q_t)(1-q_t^{(Z+1)})} \Big] \\
    &= \mathds{E}_{q_t} \Big [ \big(1-q_t^{(Z+1)}\big) \cdot \frac{\delta q_t(Zq_t^{(Z+1)} - (Z+1)q_t^Z +1)}{(1-q_t)(1-q_t^{(Z+1)})} \Big] \hspace{1cm} \mid \text{fix } 0<p<1\\
    &\geq \Big [ \underbrace{\frac{\delta p}{(1-p)}}_{>0} \cdot \underbrace{(Zp^{(Z+1)} - (Z+1)p^Z +1)}_{=: f_Z(p)} \Big] \cdot \underbrace{P(q_t = p)}_{>0}
\end{align*}
To show that $R > 0$ it remains to show that $f_Z(p) > 0$ for $ 0<p<1$. We first show that $f_Z(p)$ is strictly monotonically decreasing, $\forall 0<p<1$. Then we show that $f_Z(1)=0$. From that follows that $\forall 0<p<1:$ $f_Z(p) >0$. 

$f_Z(p)$ is strictly monotonically decreasing for $0<p<1$: 
\begin{align*}
    f'_Z(p) &= Z(Z+1)p^Z - Z(Z+1)p^{Z-1}\\
    &= Z(Z+1) (p^Z-p^{Z-1}) \\
    &< 0
\end{align*}

$f_Z(1) = Z-(Z+1) +1 = 0$ 

\noindent 
Thus, $f_Z(p) > 0$ and $R > 0.$

\subsection{Proof for Theorem \ref{th:gainpositive}}

To prove that $G(\alpha, \tau) > 0 \; \forall \alpha > 0, N > 0, \tau > r$, we differentiate between the two cases, undersupply and oversupply of labor: 

\noindent
Case 1 (undersupply of labor, $N \leq b$): 

\begin{align*}
    G(\alpha, \tau) &= u(\alpha, \tau) - u^{\text{base}} & \\
    &= \alpha u^D + (1-\alpha) u^{A} - \frac{n}{b}(r-cm) &\\
    &= \alpha \Big[ \frac{1}{\alpha N} n^D (\tau - cm) \Big] + (1-\alpha) \Big[ \frac{1}{(1-\alpha)N} n^{A} (r-cm) + \frac{n}{(1-\alpha)N} \underbrace{R}_{= 0 \text{\scriptsize{ by L. \ref{lemma:R_undersupply}}}} \Big]- \frac{n}{b}(r-cm) & \\
    &= \frac{1}{ N} \Big[ \alpha N \frac{n}{b} (\tau - cm) + (1-\alpha) N \frac{n}{b} (r-cm) \Big]- \frac{n}{b}(r-cm) & \\
    &= \frac{n}{b} \Big[ \alpha (\tau -cm) + (1-\alpha)(r-cm) - (r-cm)\Big] \\
    &= \frac{n}{b} \alpha \underbrace{(\tau - r)}_{>0} \\
    &> 0
\end{align*}

\noindent
Case 2 (oversupply of labor, $N > b$): Recall that in case of oversupply all orders get delivered and it holds $n^D+n^{A} = n$. 

\begin{align*}
    G(\alpha, \tau) &= u(\alpha, \tau) - u^{\text{base}} & \\
    &= \alpha u^D + (1-\alpha) u^{A} - \frac{n}{N}(r-cm) &\\
    &= \alpha \Big[ \frac{1}{\alpha N} n^D (\tau - cm) \Big] + (1-\alpha) \Big[ \frac{1}{(1-\alpha)N} n^{A} (r-cm) + \frac{n}{(1-\alpha)N} R \Big]- \frac{n}{N}(r-cm) & \\
    &= \frac{1}{N} \Big [ n^D (\tau - cm) + n^{A}(r-cm) + nR - n(r-cm) \Big] \\
    &= \frac{1}{N} \Big [ n^D \tau + n^{A} r - cm \underbrace{(n^D+n^{A})}_{=n} + nR - nr + ncm \Big ] \\
    &= \frac{1}{N} \Big [ n^D \tau + n^{A} r + nR - nr \Big]. \end{align*}

It follows that
\begin{align*}
    G(\alpha, \tau)= \frac{1}{N } \Big [ n^D \tau + r \underbrace{(n^{A} -n)}_{-n^D} +nR \Big]= \frac{1}{N } \Big [ n^D \underbrace{(\tau -r)}_{>0} +n\underbrace{R}_{>0} \Big ] > 0
\end{align*}

\subsection{Proof of Equation \ref{eq:Rt}} 
\label{sec:proof_eq_Rt}

We want to prove the following equation: 
\begin{equation*}
    \sum_{z=0}^{Z} \delta z\: \frac{q_t^z (1-q_t)}{(1-q_t^{(Z+1)})} = \frac{\delta q_t(Zq_t^{(Z+1)} - (Z+1)q_t^Z +1)}{(1-q_t)(1-q_t^{(Z+1)})}
\end{equation*}

\noindent
We start by rearranging the left side: 
\begin{equation*}
    \sum_{z=0}^{Z} \delta z\: \frac{q_t^z (1-q_t)}{(1-q_t^{(Z+1)})} = \frac{\delta (1-q_t)}{(1-q_t^{(Z+1)})} \underbrace{\sum_{z=0}^{Z} z q_t^z}_{:= S_Z}
\end{equation*}

\noindent
Then, we bring $S_Z$ into a closed-form solution:
\begin{align*}
    S_Z &= \sum_{z=0}^{Z} z q_t^z \\
    S_Z &= q_t + 2q_t^2 + ... + Zq_t^Z &&\mid \cdot \: q_t\\
    q_t S_Z  &= q_t^2 + 2q_t^3 + ... + Zq_t^{(Z+1)} &&\mid {\footnotesize \text{subtract $q_t S_Z$ from $S_Z$}} \\
    (1-q_t) S_Z &= [q_t + 2q_t^2 + ... + Zq_t^Z] - [q_t^2 + 2q_t^3 + ... + Zq_t^{(Z+1)}] \\
    (1-q_t) S_Z &= (q_t + q_t^2 + ... + q_t^Z) - Zq_t^{(Z+1)} &&\mid {\footnotesize \text{geom. series}} \\
    (1-q_t) S_Z &= \frac{1-q_t^{(Z+1)}}{1-q_t} -1 - Zq_t^{(Z+1)} &&\mid :(1-q_t)\\
    S_Z &= \frac{Zq_t^{(Z+2)} - (Z+1)q_t^{(Z+1)} + q_t}{(1-q_t)^2}
\end{align*}

Plugging this into the equation from above yields the desired result: 
\begin{align*}
    \sum_{z=0}^{Z} \delta z\: \frac{q_t^z (1-q_t)}{(1-q_t^{(Z+1)})} &= \frac{\delta (1-q_t)}{(1-q_t^{(Z+1)})} S_Z \\
    &= \frac{\delta (1-q_t)}{(1-q_t^{(Z+1)})} \cdot \frac{Zq_t^{(Z+2)} - (Z+1)q_t^{(Z+1)} + q_t}{(1-q_t)^2} \\
    &= \frac{\delta q_t(Zq_t^{(Z+1)} - (Z+1)q_t^Z +1)}{(1-q_t)(1-q_t^{(Z+1)})}
\end{align*}

\subsection{Proof for Theorem \ref{th:undersupplybenefitpositive}}

We want to prove that in the case of undersupply ($N\leq b$) the benefit of participation is positive and given by:
\[B(\alpha, \tau) = \frac{n}{b} (\tau - r) > 0 \]
Recall that in the case of undersupply every worker delivers $\frac{n}{b}$ orders, thus $n^D = \alpha N \frac{n}{b}$ and $n^{A} = (1-\alpha) N \frac{n}{b}$. 

\begin{align*}
    B(\alpha, \tau) &= u^D(\alpha, \tau) - u^{A}(\alpha, \tau) &\\
    &= \frac{1}{\alpha N} n^D(\tau - cm) - \Big[ \frac{1}{(1-\alpha) N} n^{A} \big(r - cm\big)  + \frac{n}{(1-\alpha) N} \underbrace{R}_{=0 \text{\tiny{ by L. \ref{lemma:R_undersupply}}}} \Big ]\\
    &= \frac{1}{\alpha N} \cdot \alpha N \frac{n}{b}(\tau - cm) -\frac{1}{(1-\alpha) N} \cdot (1-\alpha) N \frac{n}{b} \big(r - cm\big)  \\
    &= \frac{n}{b} (\tau -cm) - \frac{n}{b} (r-cm) \\
    &=\frac{n}{b} (\tau - r) > 0
\end{align*}

\subsection{Proof for Theorem \ref{th:oversupplygroupsizethreshold}}

Assume that $r>cm$. We want to prove that in case of oversupply, the following condition on degree of oversupply $\deg$, group size $\alpha<1$ and threshold $\tau$ is necessary such that the benefit of participation is positive: 
\[ \deg < \frac{1}{1-\alpha} \Big ( 1- \frac{\alpha (r+R-cm)}{(1-\alpha)(\tau - cm) + \alpha (r-cm)} \Big ) \Rightarrow B(\alpha, \tau) > 0 \] 

\noindent
First, recall the oversupply lower bound on $\gamma_D$ from Lemma \ref{lemma:undersupply_nb}: 
\begin{align*}
        \gamma_D \geq \textnormal{max } \Big(0, 1-(1-\alpha) \cdot \deg\Big)
\end{align*}

\noindent
We first plug in the definition of $\gamma_D$ which in the case of oversupply simplifies to $\gamma_D = \frac{n^D}{n}$. Then we plug in the bound and denote the result by $(\star)$:

\begin{align*}
    B(\alpha, \tau) &= u^D - u^{A} \\
    &= \frac{1}{\alpha N} n^D(\tau - cm) - \Big[ \frac{1}{(1-\alpha)N} n^{A} (r -cm)  + \frac{n}{(1-\alpha)N} R \Big ]\\
    &= \frac{1}{\alpha N} n \gamma_D (\tau - cm) - \frac{1}{(1-\alpha)N} n(1-\gamma_D) (r -cm)  - \frac{n}{(1-\alpha)N} R \Big ]\\
    &= n \gamma_D \Big [ \frac{\tau -cm}{\alpha N} + \frac{r-cm}{(1-\alpha)N} \Big] - \frac{n}{(1-\alpha)N} \big( r-cm + R \big ) \\
    &\geq \underbrace{\textnormal{max} \Big(0, 1-(1-\alpha) \cdot \deg\Big) n \Big[ \frac{\tau -cm}{\alpha N} + \frac{r-cm}{(1-\alpha)N} \Big] - \frac{n}{(1-\alpha)N} \big( r-cm + R \big ) }_{= (\star)}
\end{align*}

At this point we distinguish between the two cases to which max$(\cdot, \cdot)$ can evaluate: 

\noindent
Case 1: $\textnormal{max } \Big(0, 1-(1-\alpha) \cdot \deg\Big) = 0$, then 
\[(\star ) = 0 - \frac{n}{(1-\alpha)N} \big( r-cm + R \big ) \]
which is negative given $r > cm$. Therefore the second case is the interesting one: 

\noindent
Case 2: $\textnormal{max } \Big(0, 1-(1-\alpha) \cdot \deg\Big) = 1-(1-\alpha) \cdot \deg$, then:  \\

First, we slightly transform the condition from the theorem: 
\begin{align*}
    \deg &< \frac{1}{1-\alpha} \Big ( 1- \frac{\alpha (r+R-cm)}{(1-\alpha)(\tau - cm) + \alpha (r-cm)} \Big ) \\
    -\deg &> \frac{1}{1-\alpha} \Big ( \frac{\alpha (r+R-cm)}{(1-\alpha)(\tau - cm) + \alpha (r-cm)} -1 \Big )
\end{align*}
Then we return to $(\star)$, the lower bound on $B(\alpha, \tau)$, and use the condition from the theorem to show that $B > 0$: 

\begin{align*}
    (\star ) &= (1-(1-\alpha) \cdot \deg) n \Big[ \frac{\tau -cm}{\alpha N} + \frac{r-cm}{(1-\alpha)N} \Big] - \frac{n}{(1-\alpha)N} \big( r-cm + R \big )  &\\
    &= n \Big [ \frac{(1-\alpha)(\tau-cm) + \alpha (r-cm)}{(1-\alpha)\alpha N} - (1-\alpha) \deg \frac{(1-\alpha)(\tau-cm) + \alpha (r-cm)}{(1-\alpha)\alpha N} \\&\qquad \qquad\qquad- \frac{\alpha (r-cm+R)}{(1-\alpha)\alpha N} \Big] \\
    &> n \Big [  \Big ( \frac{\alpha (r+R-cm)}{(1-\alpha)(\tau - cm) + \alpha (r-cm)} -1 \Big ) \cdot \frac{(1-\alpha)(\tau-cm) + \alpha (r-cm)}{(1-\alpha)\alpha N}\\&\qquad\qquad\qquad + \frac{(1-\alpha)(\tau-cm) - \alpha R }{(1-\alpha)\alpha N} \Big ] \\
    &= n\Big [ \frac{(1-\alpha)(\tau-cm) - \alpha R + \alpha(r+R -cm) - (1-\alpha)(\tau-cm) - \alpha (r-cm)}{(1-\alpha)\alpha N} \Big] \\
    &= 0
\end{align*}
This shows that $B(\alpha, \tau) >0$ if $\deg < \frac{1}{1-\alpha} \Big ( 1- \frac{\alpha (r+R-cm)}{(1-\alpha)(\tau - cm) + \alpha (r-cm)} \Big )$.

\subsection{Proof for Theorem \ref{th:freeriding}}

We want to prove that in the case of undersupply the benefit of freeriding is 0, $F(\alpha, \tau) = 0$. Recall that in the case of undersupply every worker delivers $\frac{n}{b}$ orders, thus $n^D = \alpha N \frac{n}{b}$ and $n^{A} = (1-\alpha) N \frac{n}{b}$.

\begin{align*}
    F(\alpha, \tau) &= u^A(\alpha, \tau) - u^{\text{base}} & \text{\tiny{(Eq. \ref{eq:u0})}}\\
    &= \frac{1}{(1-\alpha)N}n^A \cdot (r-cm) + \frac{n}{(1-\alpha)N} R - \frac{n}{b}(r-cm) & \\
    &= \frac{n}{b} (r+\underbrace{R}_{=0 \text{\tiny{ by L. \ref{lemma:R_undersupply}}}} - cm - (r-cm)) & \\
    &= 0
\end{align*}

\section{Participant fraction of delivered orders and spillover effect}

\begin{figure}[ht!]
\includegraphics[width=0.9\textwidth]{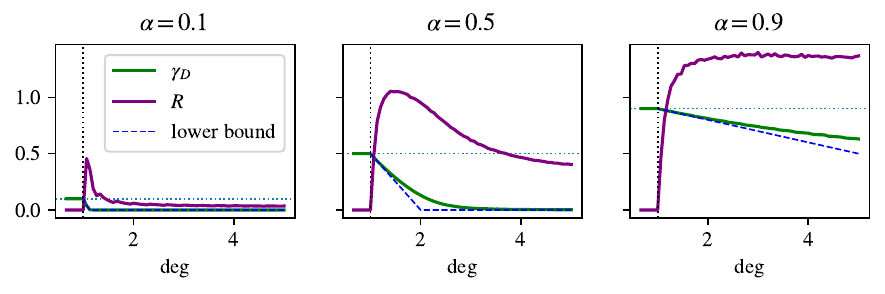}
\centering
\caption{The participant fraction of delivered orders $\gamma_D$ (green), the lower bound on $\gamma_D$ (blue dashed) and the spillover effect $R$ for $b=30$ and $\alpha \in \{0.1, 0.5, 0.9\}$}
\label{fig:gamma_R}
\end{figure}

Figure \ref{fig:gamma_R} shows the behaviour of the participant fraction of delivered orders $\gamma_D$ and the spillover effect $R$ as a function of the degree of oversupply, as well as the lower bound on $\gamma_D$ obtained in Lemma \ref{lemma:undersupply_nb}. This illustrates the results from Lemma \ref{lemma:undersupply_nb} and Lemma \ref{lemma:R_undersupply}.

\section{Additional figure for Section \ref{sec:workinginshifts}}
\label{sec:additionalfigureshifts}

\begin{figure}[h!]
\centering
\includegraphics[width=0.43\linewidth]{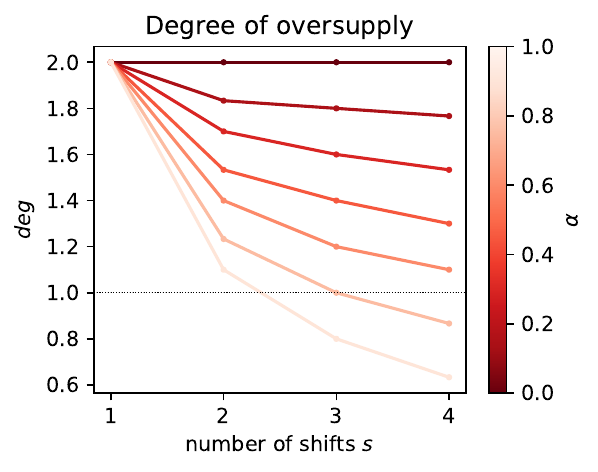}
  \label{fig:shifts_utility_singleN}
  \caption{\textbf{(left)} Benefit of participation $B$ for different number of shifts, $s=1, .., 4$. Compare to the middle scenarion from Figure \ref{fig:shiftstrategy_N}. $B$ increases with the number of shifts. \textbf{(right)} Working in shifts to counter undersupply. Degree of oversupply, deg, decreases with number of shifts $s=1, .., 4$, depicted for different level of $\alpha$ and fixed $N=60$.}
\label{fig:shiftstrategy_alpha}
\end{figure}

Working in multiple shifts decreases the degree of oversupply as shown in Figure \ref{fig:shiftstrategy_alpha}. For each initial participant level $\alpha$ we compute how the degree of oversupply changes, by computing the new number of all drivers $\widetilde{N} = N- (\alpha - \frac \alpha s)N$. 

\end{document}